\documentclass[sigconf]{acmart}
\newcommand{\eat}[1]{}

\usepackage{xspace}
\usepackage{multirow}
\usepackage{balance}
\usepackage{ctable}
\usepackage{listings}
\usepackage{color}
\usepackage{graphicx}
\usepackage{subcaption}
\usepackage{syntax}

\usepackage{multirow}

\usepackage{algorithm}
\usepackage{algpseudocode}

\usepackage{amssymb}
\usepackage{amsmath}
\usepackage{amsfonts}
\usepackage{hyperref}

\usepackage{cleveref}
\usepackage{acronym}
\usepackage{fancybox}
\usepackage{verbatim}
\usepackage[normalem]{ulem}

\usepackage{newfloat}
\usepackage{mdframed}
\usepackage{enumitem}

\definecolor{red}{RGB}{255,0,0}

\newlength{\MaxSizeOfLineNumbers}%
\definecolor{keywordcolor}{rgb}{0.8,0.1,0.5}
\definecolor{lightlightgray}{gray}{.96}
\definecolor{lightgray}{gray}{.925}
\definecolor{medlightgray}{gray}{0.7}
\definecolor{medgray}{gray}{0.4}
\definecolor{darkgray}{gray}{0.35}
\definecolor{nearblack}{gray}{0.15}

\crefname{component}{Component}{Components}
\newcommand{\oneColFig}[2]{
\begin{figure}[htbp]
\centering
\includegraphics[width=0.45\textwidth]{figs/#1}
\caption{#2}
\label{fig:#1}
\end{figure}
}
\newcommand{\fullColFig}[2]{
\begin{figure*}[htbp]
\centering
\includegraphics[width=0.9\textwidth]{figs/#1}
\caption{#2}
\label{fig:#1}
\end{figure*}
}
\newcommand{\prgref}[1]{Program~\ref{prg:#1}}

\definecolor{eclipseBlue}{RGB}{42,0.0,255}
\definecolor{eclipseGreen}{RGB}{63,127,95}
\definecolor{eclipsePurple}{RGB}{127,0,85}

\floatstyle{plain}

\newfloat{Program}{h}{eqn}

\definecolor{MyDarkGreen}{rgb}{0.0,0.4,0.0} % This is the color used for comments
\newcommand{\javafile}[2]{
\begin{Program}
\lstinputlisting[]{code/#1.java}
\caption{#2}
\label{prg:#1}
\end{Program}
}
\newcommand{\xmlfile}[2]{
\begin{Program}
\lstinputlisting[]{code/#1.xml}
\caption{#2}
\label{prg:#1}
\end{Program}
}

\newcounter{rqs}
\acrodef{rq}[RQ]{research question}
\acrodef{ui}[UI]{user interface}
\acrodef{mwu}[MWU]{Mann Whitney U}
\acrodef{uiat}[UIAT]{\textit{UIAutomator}}
\acrodef{man}[MAN]{Mobile Ad Network}
\acrodef{tdl}[TDL]{Tracking Description Language}
\acrodef{apt}[APT]{Advanced Persistent Threat }

\acrodef{dsl}[DSL]{Domain Specific Language }
\acrodef{nn}[CNN]{Convolutional Neural Network }
\newcommand{\prgname}{Sextant~}
\acrodef{ast}[AST]{Abstract Syntax Tree }
\acrodef{cfg}[CFG]{Control Flow Graph }

\DeclareFloatingEnvironment[
  fileext   = logr,% don't use log here!
  listname  = {List of Grammars},
  name      = Grammar,
  placement = htp
]{Grammar}

\definecolor{mygreen}{rgb}{0,0.6,0}

\setcopyright{none}
\begin{document}

\title{ Detecting Low Rating Android Apps Before They Have Reached the Market}

\author{Ding Li}
\affiliation{%
  \institution{NEC Labs America, Inc.}
  %\streetaddress{P.O. Box 1212}
  \city{Princeton}
  \state{New Jersey}
  \postcode{08540}
}
\email{dinglipersonal@outlook.com}

\author{Dongjin Song}
\affiliation{%
  \institution{NEC Labs America, Inc.}
  %\streetaddress{P.O. Box 1212}
  \city{Princeton}
  \state{New Jersey}
  \postcode{08540}
}
\email{songdj2008@gmail.com}

\begin{abstract}
%  Driven by the popularity of the Android system, Android app markets enjoy a booming prosperity. One critical problem for modern Android app markets is to detect and prevent apps that are going to receive low ratings from end users. To do this, current approach has to publish an app first and collect enough user ratings and reviews before. Then, current approach  can detect if the app is appreciated by end users or not. However, by doing so, the reputation of app markets has already been damaged.
  
Driven by the popularity of the Android system, Android app markets enjoy a booming prosperity in recent years. One critical problem for modern Android app markets is how to prevent apps that are going to receive low ratings from reaching end users. For this purpose, traditional approaches have to publish an app first and then collect enough user ratings and reviews so as to determine whether the app is favored by end users or not. In this way, however, the reputation of the app market has already been damaged.
 
%In this paper, we propose a novel technique to detect the low rating Android apps before they have reached any end users. With our technique, an Android app market does not need to risk it reputation to expose low rating apps to users. Our approach is based on novel static analysis techniques and machine learning techniques. As we evaluated, our approach can achieve on average 90.50\% precision and 94.31\% recall.

To address this problem, we propose a novel technique, \textit{i.e.}, \prgname, to detect low rating Android apps based on the .apk files. With our proposed technique, an Android app market can prevent from risking it reputation on exposing low rating apps to users. \prgname is developed based on novel static analysis techniques as well as machine learning techniques. In our study, our proposed approach can achieve on average 90.50\% precision and 94.31\% recall.

\end{abstract}
% \keywords{ACM proceedings, \LaTeX, text tagging}

\maketitle
\thispagestyle{empty}
%\begin{keywords}

%\end{keywords}

%\category{H.4}{Information Systems Applications}{Miscellaneous}
%\category{D.2.8}{Software Engineering}{Metrics}[complexity measures, performance measures]
%\terms{Theory}
\acresetall

\section{Introduction}
\label{sec:introduction}
Nowadays, Android is the most popular mobile platform. 88\% of mobile phones are running
Android~\cite{androidstat}. Such a popularity has stimulated a prosperity in the Android
app market. According to one of the recent reports, the total revenue of Android apps has
reached 27 billion US dollars in 2016~\cite{apprevenue}. Such a prosperity does not only
exist in the official Google Play market but also in many other third party markets, which
produce 10 billion US dollars revenue in 2016~\cite{apprevenue}. Today, millions of apps
could be downloaded in Android markets. Thousands of apps are created and uploaded to those markets~\cite{androidnumber} everyday.

Despite the prosperity of Android markets, many apps in the
market often receive low star ratings. These apps are not appreciated by end users. They either provide inferior user experience or have low code quality. Having too many low rating
apps in an app market can hurt the reputation of the market. It will eventually drive end users away
from the market and make the market suffer from losing revenue.
This problem can be even more severe for third party markets since they have more competitions than the Google Play market.

Due to this reason, the app markets are trying to prevent low rating
apps.  Google Play is working on punishing the
developers of the apps with lowest star ratings~\cite{lowqualitypunishment}.
This approach is useful, but it is more about fixing the damages rather than preventing
the damages. Currently, the only way for an app market to know whether an app will have a low rating is to publish the app first, accumulate
star ratings from a substantial amount of end users, and finally average these star ratings. In such a process,
the low rating app has already reached the end users. It means that the low rating app has already caused damages to the reputation of the market.
Another approach to prevent low rating apps is to have manual inspections~\cite{ios}. However, this approach is labor
intensive and expensive.

Therefore, it is beneficial for app markets to be able to automatically detect low rating
apps without any user feedback. It benefits an app market in two ways. First, the market does not
need publish low rating apps in order to detect them. When a potential low rating app is uploaded, it will
be automatically detected and prevented from the market. In this way, the reputation of the market will not be damaged.
Second, the automated detection technique can be used to assist the manual inspection and accelerate the
review process. This could save the labor and expense in the manual inspection process.

Being able to detect low rating apps can also be very beneficial for developers and end users.
For developers, such a capability enables them to have a quick feedback about their app without waiting for a few weeks for actual user ratings. If they find that
their app may potentially be a low rating app, they could have an early plan for modification.
For end users, this capability can be useful when users are installing apps from unknown sources.
For these apps, they rarely have valid user ratings. Thus, it is very difficult
for end users to know the quality of an app from unknown sources.  Being able to detect low rating apps can
potentially help end users avoid wasting their time on the apps that they do not expect.

Although automatically detecting low rating apps is very valuable, it is very challenging to achieve such a goal.
The main challenge comes from the fact that app ratings are highly subjective and abstract. App ratings are subjective because they are provided
by end users. Thus, they are inevitably affected by the personal preference of each end user. Such a personal preference is
very difficult to be modeled. App ratings are abstract because they describe the general feeling of end users to an app. There isn't a concrete rule or algorithm to generate the rating of an app. Thus, it is impossible to detect low rating apps by detecting a pre-defined code pattern or bug.

% \textbf{[App ratings are general because they are not designed for any certain feature or aspect of an app]?}.
% Instead, app ratings are designed to measure the general feelings of users for the app. 

Traditional program analysis techniques are not capable of detecting low rating Android applications.
Traditional static or dynamic
program analysis techniques work well on detecting and fixing specific problems in
programs~\cite{7886908,Zhang:2015:TAG:2810103.2813669,Zhang:2016:PMT:2931037.2931038,7194604} or
modeling concrete metrics, such as
energy~\cite{6606555,Li:2013:CSL:2483760.2483780,Liu:2014:CDP:2568225.2568229} or
runtime~\cite{Xu:2010:FLD:1806596.1806617,Xiao:2013:CDI:2483760.2483784,Nistor:2013:TDP:2486788.2486862}. However, since the rating of an Android app is subjective and abstract, detecting the specific code problems or modeling specific metrics is not sufficient to detect low rating apps.

To automatically detect low rating Android apps, in this paper, we propose a static analysis and machine learning based approach, Sextant.  Our approach detects if an Android app is low rating only based on the .apk package of the app. Our implementation of \prgname is available on github\footnote{https://github.com/marapapman/Sextant}. To the best of our knowledge, this is the very first technique to do the similar task.

\prgname contains three main components. First, it contains two novel representations for an Android app. These two representations can be 
retrieved from the .apk file of an app with effective and scalable static analysis techniques. The first representation is the semantic vector,
which is used to capture the features of the executable file of the app. The second one is the layout vector, which is used to represent the layout information of the app. Second, \prgname contains two pre-training neural network models to learn unified features from both executable files and UI layout of an Android app. Third, \prgname contains a neural network model that accepts features from the pre-training model and determines whether an Android app is low rating or not.

We also perform an extensive empirical study with \prgname. Specifically, we measure the detection accuracy of \prgname on 33,456 realistic Android apps from the Google Play market. In our experiment, \prgname on average achieved 92.31\% accuracy, 90.50\% precision, and 94.31\% recall. We also compare \prgname with four other baseline models:  two models which use the bag of words representation~\cite{Peiravian2013ICTAI,sahs2012EISIC,Shamili2010ICPR,Aafer2013,schmidt2009static,burguera2011crowdroid,enck2009lightweight,zhou2012hey,arp2014drebin}, the executable only model, and the UI only model. The proposed \prgname outperforms all the four baseline models with statistically significant differences. 

This paper has following contributions
\begin{itemize}
  \item Our approach is the very first approach to detect low rating Android apps before they can reach the end users.
  \item We propose a novel representation to model the semantics of Java/Android apps
  \item Extensive evaluation of the proposed approach on 33,456 Google Play apps demonstrates the effectiveness of the proposed method.
  
\end{itemize}

The other parts of this paper are organized as follows. In \Cref{sec:background}, we briefly discuss the background information of Android apps and convolutional neural networks. In~\Cref{sec:approach}, we discuss the approach of \prgname. In~\Cref{sec:preliminary}, we discuss a preliminary study to evaluate how accurate can our executable representation capture the differences and similarities between Android/Java applications on the semantic level. In~\Cref{sec:evaluation}, we discuss the evaluation result of \prgname. In~\Cref{sec:threat}, we discuss the threat to validity of our evaluation. In~\Cref{sec:related}, we discuss the related work. Finally, we conclude this paper in~\Cref{sec:conclusion}. 

% Machine Learning techniques are proven to be useful to model subjective and general metrics. They are widely used to classify malwares~\cite{Zhang:2014:SAM:2660267.2660359,} and detect files that contain defects~\cite{}.

\section{Background}
\label{sec:background}
In this section, we briefly discuss some background
information about the structures of Android apps and the principle of
convolutional neural networks.
\subsection{Structures of Android Apps}
An Android app is organized as a .apk package, which includes all the resources and
code for the app. Among those resources, the executable is organized as the .dex file
and the UI layout files are stored as XML files in the layout/ folder.

The executable of an Android app is in the form of Dalvik bytecode, which is compiled
from Java. Therefore, many tools, such as soot~\cite{soot} and dex2jar~\cite{dex2jar}
can be used to convert the Dalvik bytecode to Java bytecode. The executable of an
Android app contains several activities, which are basic components of the Android
app. Each activity starts at the ``onCreate'' callback~\cite{androidcode}. 

The UI layout XML files define the visual structure for user interfaces. They can be
loaded in the ``onCreate'' callback of an activity to create the GUI. The basic blocks of
 the UI layout XML files are the \textit{View} and \textit{ViewGroup} tags. A
 \textit{View} tag represents the basic UI elements such as text boxes, buttons, and graphs. A \textit{ViewGroup} tag is a special type of  \textit{View}, it represents a
group of other \textit{View} tags. Combining the \textit{View} and  \textit{ViewGroup} tags, developers can declare the UI of an app as a  layout tree~\cite{androidUI}.
\subsection{Convolutional Neural Networks}

%Convolutional Neural Network is a popular of deep neural network model in machine learning. It simulates the biological process that each neutrons is only activated by a restricted area of the input. Compared to normal neural network structure, the convolutional neural network has a convolutional layer, which moves a sliding window along the input with a fixed step length. While moving, the sliding window updates the weight of the connections and learns features about the subset of the input in the sliding window. By doing this, convolutional neural network can reduce the weights that need to be learned and can learn features about structured information.

Convolutional Neural Networks (CNNs) are a popular deep neural network model which has been widely utilized for image classification, machine translation, \textit{etc}. It simulates the biological process that each neutron is only activated by a restricted area of the input. Compared to ordinary neural network structures such as restricted boltzmann machine or autoencoder, typical convolutional neural networks are comprised of one or more convolutional layers (often with a subsampling step) and then followed by one or more fully connected layers as in a standard multilayer neural network. The convolutional layer is the key building block of a convolutional neural network. During the forward pass, each filter is slided across the width and height of the input volume to compute dot product between the entries of the filter and the input at any position. In this way, convolutional neural networks can reduce the weights that need to be learned and obtain meaningful representations of the input structured information.
\oneColFig{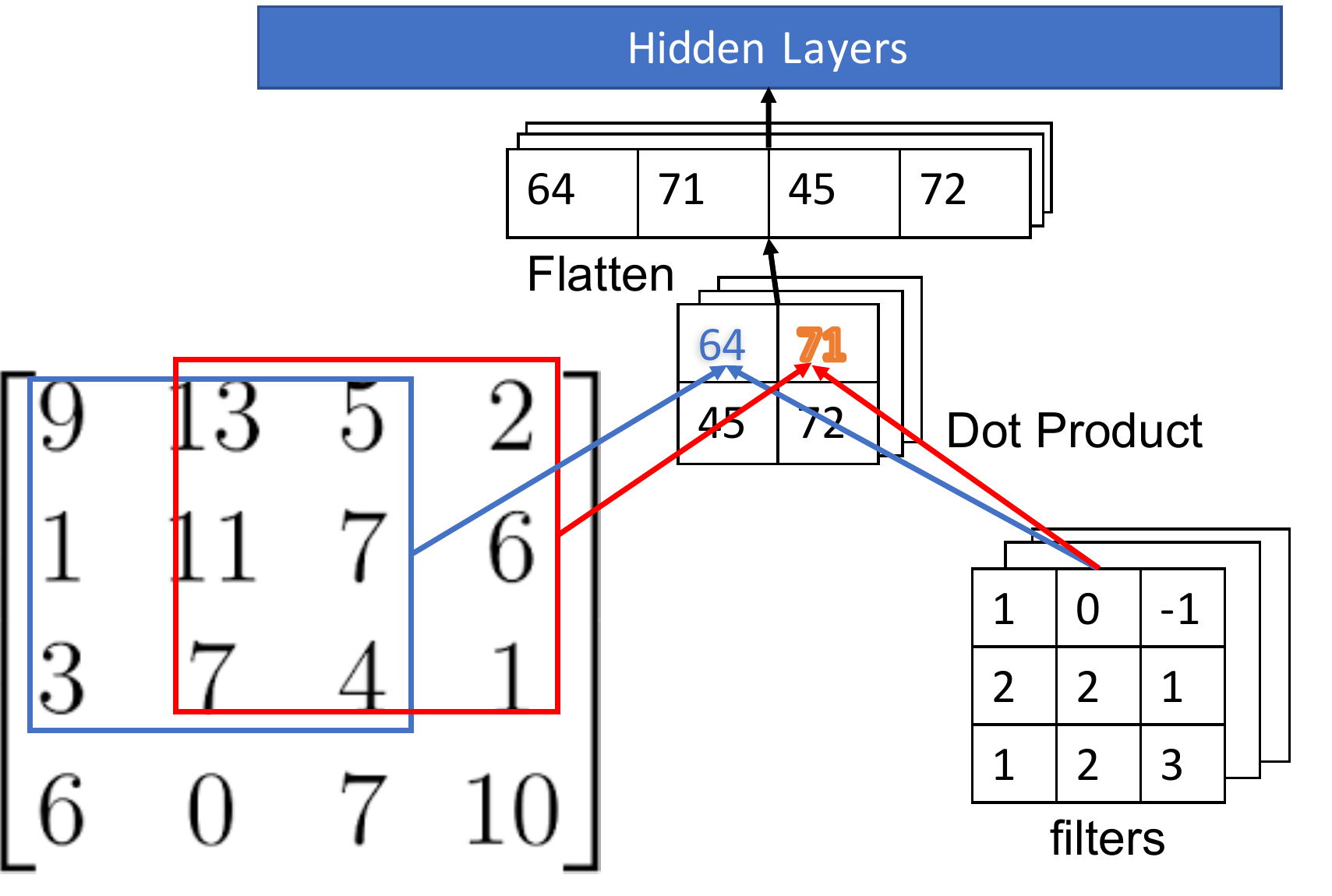}{An example of convolutional neural networks}
\Cref{fig:CNN} shows an example of convolutional neural networks. In this figure, the input is a 4$\times$4 matrix. The sliding window, which has a size of 3$\times$3, is shown as the blue and red boxes in \Cref{fig:CNN}. Each color represents a step of moving the sliding window. In a forward pass, the convolutional layer first computes dot product between the 3$\times$3 filter and the 3$\times$3 sliding window on the input, then it moves the sliding window with stride 1 to the next 3$\times$3 input entries and computes its dot product with the 3$\times$3 filter again. This process is repeated until all elements in the input matrix are processed. After the convolution operation with a number of filters, multiple feature maps are produced, subsampled, and transformed into a hidden feature vector (hidden layer) which can be eventually utilized for classification or other tasks.

\section{Approach}
\label{sec:approach}
\fullColFig{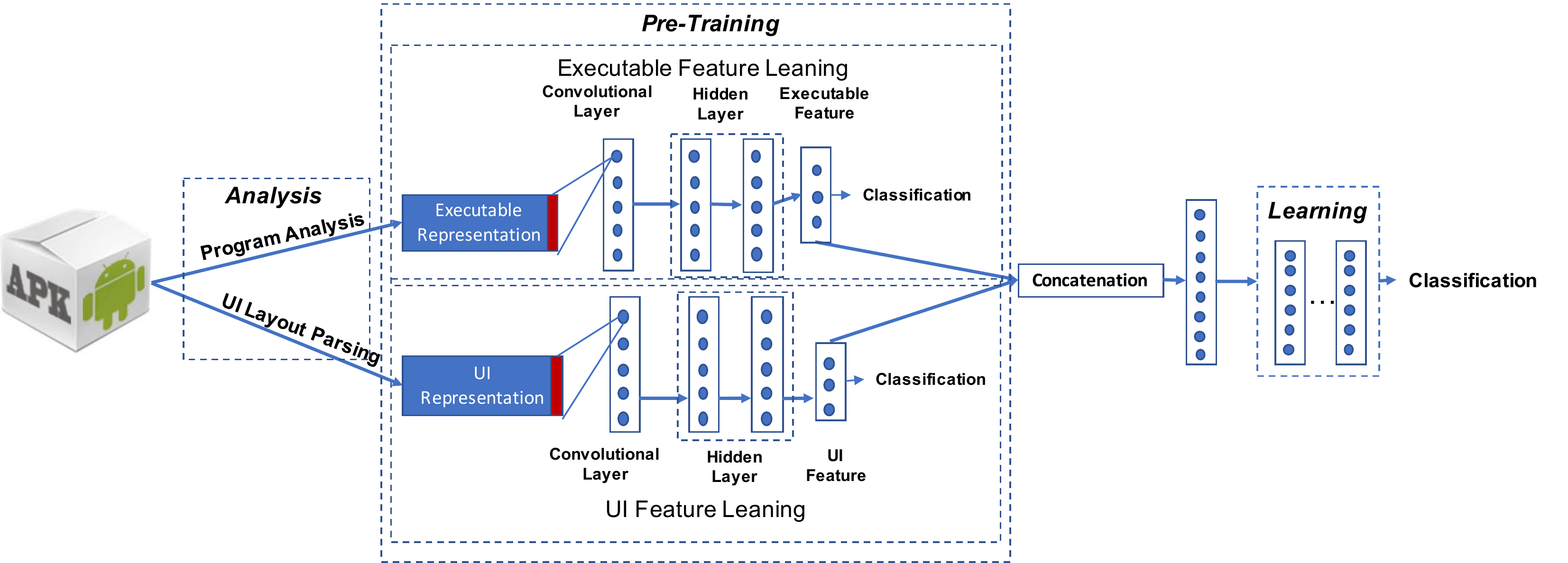}{The workflow of our approach}

% \begin{figure*}
% %\vspace{-0.1cm}
% \centering
% \includegraphics[width=0.90\textwidth]{figs/Flowchart.pdf}
% %\caption{Diagram for attach reproduction scenario}
% \vspace*{-1ex}
% \caption{The workflow of our approach}
% \label{fig:Flowchart}
% \vspace*{-4ex}
% \end{figure*}
Our approach takes the .apk of an Android app as input and outputs whether the input app is a low rating app. The workflow of our approach is
shown in \Cref{fig:Flowchart}. It has three stages. The first stage is the
analysis stage. In this stage,
\prgname  obtains the executable
representation, \textit{i.e.}, the semantic vector, and the UI representation, \textit{i.e.}, the layout vector of the input app respectively.
The executable representation is obtained by using static program analysis techniques.
The UI representation is obtained by parsing the UI layout files of the app. The
second stage is the pre-training stage. In this stage, \prgname feeds the executable
representation and the UI representation to two \ac{nn} models respectively.
This stage learns the normalized features of the executable and the UI. In the
final stage, \textit{i.e.}, the learning stage, \prgname concatenates the executable feature and the
UI feature, which are obtained in the pre-training stage, as a global feature vector and then feeds it to a multilayer perceptron followed by a softmax function to perform the low rating app detection. The detection process of the final stage is essentially a classification process. The softmax function determines whether the input app belongs to the low rating app or not.

%% dongjin comment: NN-> CNN ?

\subsection{Executable Representation}
\label{sec:exerepresentation}
In \prgname, the executable representation of an Android app is the semantic vector, which is
defined as $\langle I_1, I_2 ... I_N \rangle$, where $N$ is the total types of instructions in the
Android framework. Here, instructions include all basic operations, such as
arithmetic add and branching instructions, and all the APIs in the Android framework,
such as file operations or network communications. All types of instructions are numbered
from 1 to $N$. $I_k$ is the feature of the $kth$ type of instructions. It is a 3-tuple $(f_k, l_k, b_k)$, where $f_k$ is the frequency of instructions with
 type $k$ in the input app, $l_k$ is average loop depth of each instruction with type $k$, and $b_k$ is the average number of branches that instructions with type $k$ are contained by. For the rest parts of this paper, the terms semantic vector and executable representation are interchangeable.

 \javafile{ExecutableRepresentation}{Example code for the executable representation}

 We use the \prgref{ExecutableRepresentation} as an example to explain the semantic vector. In this example, if we neglect reference
instructions and jump instructions, the program has six
types of instructions. These instructions are listed
 in~\Cref{tbl:iddescription}

\begin{figure}
%\vspace{-0.1cm}
\centering
\includegraphics[width=0.40\textwidth]{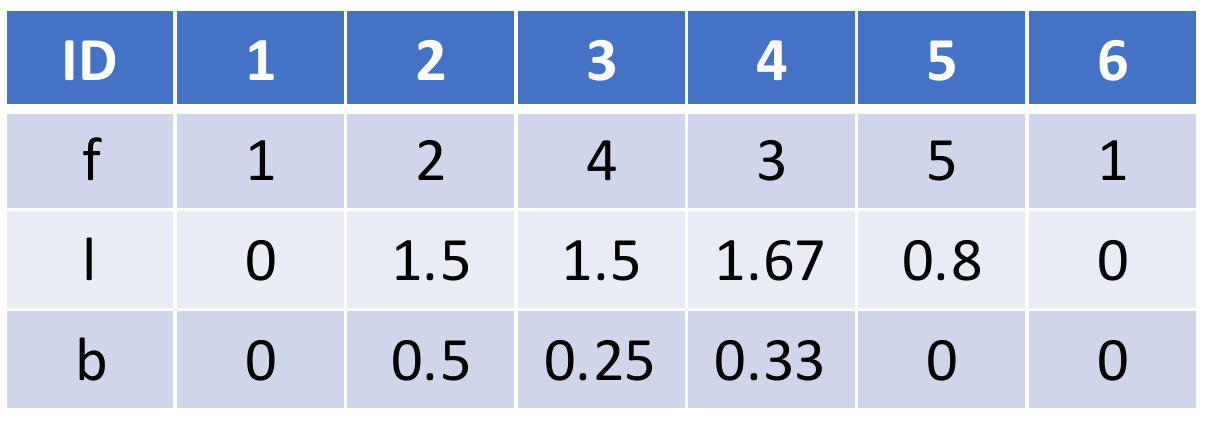}
\vspace{-1ex}
\caption{The semantic vector of \prgref{ExecutableRepresentation}}
\label{fig:ExecutableVector}
%\vspace{-4ex}
\end{figure}

\begin{table}
  \centering
  \begin{tabular}{|c|c|}
    \hline
 ID& Description \\
  \hline
  1 & The new instruction at line 2\\\hline
  2 & The $list.add$ operation at line 4 and 7\\\hline
  3 & The arithmetic add operation at line 3, 4, 5, and 7\\\hline
  4 & The comparison instruction at line 3, 5, and 6\\\hline
  5 & The assignment instruction at line 2, 3, and 5\\\hline
  6 & The $system.out.println$ at line 11\\\hline

  \hline
  \end{tabular}
  \caption{The ID of each type if instructions in \prgref{ExecutableRepresentation}}

  \label{tbl:iddescription}

\end{table}
The semantic vector of \prgref{ExecutableRepresentation} is shown in \Cref{fig:ExecutableVector}.  Each column of \Cref{fig:ExecutableVector} is 
the 3-tuple of one instruction. The ID of each column in \Cref{fig:ExecutableVector} is the
same as in the~\Cref{tbl:iddescription}. The rows $f$, $l$, and $b$ represent the frequency, average loop depth, and average branch count respectively. Among all the instructions, we use the arithmetic add operation as an
example, which is the column with ID 3 in~\Cref{fig:ExecutableVector}. The arithmetic add operation
appears four times in the program, so, the total number $f_3$ is 4. The i++ and i+1
at line 3 and 4 are in the outer loop, so their loop depths are 1. The j++ and j+i at
line 5 and line 7 are in the inner loop, so their loop depths are 2. Thus, the average
loop depth of the arithmetic add operation, \textit{i.e.}, $l_3$, is $\frac{1+1+2+2}{4}=1.5$. Only the j+i at line 7 is in a branch. Thus, the average branch count of the arithmetic add operation, \textit{i.e.}, $b_3$ is $\frac{0+0+0+1}{4}=0.25$.

\subsection{UI Layout Representation}
\prgname uses the layout vector to represent the UI layout of the target app.
Similar to the semantic vector, the layout vector is a vector of tuples. Formally, a
layout vector is defined as $\langle U_1, U_2 ... U_{M+2} \rangle$, where $M$ is the
total types of UI elements in the Android framework.
$U_{k}, 1<=k<=M$ represents the the $kth$ type of UI elements. 
$U_{M+1}$ represents the UI
elements that are from the Android Legacy Library. $U_{M+2}$ represents the customized UI elements from developers. Each $U_k$ in the layout vector is a 2-tuple $(n_k, d_k)$, where  $n_k$ and $d_k$ are the frequency and the average depth in the layout tree of the UI elements with the $kth$ type respectively. 
 For the rest parts of this paper, the terms layout vector and UI representation are interchangeable.

\xmlfile{UIlayout}{The example UI layout}

\begin{figure}
%\vspace{-0.1cm}
\centering
\includegraphics[width=0.28\textwidth]{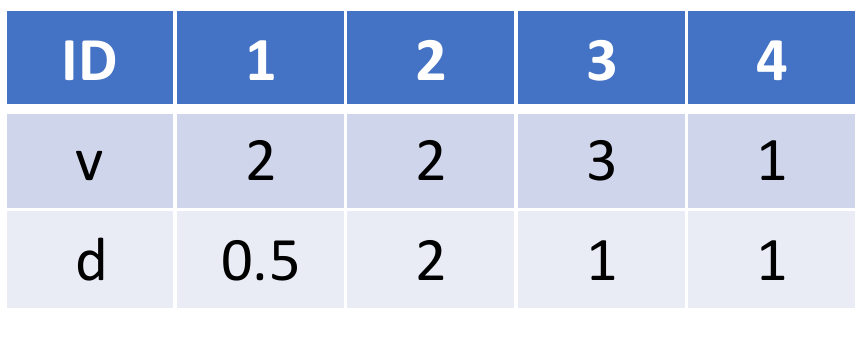}
\vspace{-1ex}
\caption{The layout vector of \prgref{UIlayout}}
\label{fig:LayoutVector}
%\vspace{-4ex}
\end{figure}

An example of UI layout file is shown in~\prgref{UIlayout}, which is one piece of layout files that are retrieved from the market app, Facebook~\cite{facebook}. This example uses two UI elements
from the Android framework, ``LinearLayout'' and ``TextView''. It uses two UI elements
from the legacy library, whose tag names start with ``android.support''. It also contains a
customized UI element, which is ``com.facebook.resources.ui.FbTextView''. To encode
this layout file as a layout vector. Our approach first numbers the UI elements. Our approach sets the IDs of ``LinearLayout'' and ``TextView'' as 1 and 2 respectively. In this
case, the UI elements from the legacy library have the ID of 3 and customized
elements have the ID of 4. Since one of the ``LinearLayout'' tags has the depth of zero in the XML tree and
another one has the depth of one. Hence, the average depth of ``LinearLayout'' is 0.5.
Similarly, the average depth of other UI elements can be calculated and the layout vector of \prgref{UIlayout} would be \Cref{fig:LayoutVector}.

\subsection{The Analysis Stage}
\label{sec:representation}
The input of the analysis stage of \prgname is the .apk file of an Android app. The
output of it is the semantic vector as the executable representation and the layout
vector as the UI layout representation. To do so, \prgname first unpacks the input
Android app to retrieve the binary executable file and the UI layout XML files of the app. Then it analyzes the executable file to calculate the frequency, average loop depth,
and average branch count of the semantic vector. It also parses the layout files to generate the
layout vector. In this section, we will focus on how to generate the semantic vector
from the executable file because the process of getting the layout vector from the layout files is similar but more straightforward.

We first introduce the intra-procedural analysis of getting the semantic vector. This process is shown in~\Cref{alg:intra}, which
takes the binary code of a method and generates the semantic vector of the method.
\Cref{alg:intra} first builds the nested loop tree and detects all branches of the
method. Then it parses all the instructions in the method and updates the loop depth and the branch counts accordingly for each instruction. At line 3 to 8, \Cref{alg:intra} first calculates the frequency, the total loop depth, and the branch count for each type of instructions. Then it averages the loop depth and the branch count by
dividing the total loop depth and branch count over the frequency.

\Cref{alg:intra} works directly on the binary code of a method. Thus, it does not have the \ac{ast} to identify the boundaries of loops and branches. To detect loops, \prgname uses the standard algorithm~\cite{aho06compiler}. For branches, instead of
detecting the branch structures directly, \prgname detect branches of the \ac{cfg} of the method.

In \prgname, a branch is defined as a 3-tuple $\langle s, e, I
\rangle$ where $s \not \in I$ and $e \not \in I$. In this tuple, $s$ is the starting point of the branch. It is a branch
instruction, such as ``if'',  and is not the source of a back edge. $e$ is the immediate
post-dominator of $s$. $I$ is all the instructions between $s$ and $e$. In \Cref{alg:intra}, $BuildBranchSet()$ first removes all the back edges of $M$ and then finds all branches in the
method.  $BranchTree.getCount(i)$ counts how many branches that contain the instruction $i$.

\begin{algorithm}
  \caption{Intra-procedural semantic vector building}
\label{alg:intra}
  \begin{algorithmic}[1]
  \Require{A method $M$ and the ID map $MAP$}
  \Ensure{The semantic vector $V$}
    \State{$LoopTree \leftarrow BuildNestedLoopTree(M)$}
    \State{$BranchSet \leftarrow BuildBranchSet(M)$}
     \ForAll{$Instruction\ i \in M$}
       \State{$ID \leftarrow MAP.getID(i)$}
       \State{$V[ID].f\mathrel{+}=1$}
       \State{$V[ID].l\mathrel{+}=LoopTree.getDepth(i)$}
      \State{$V[ID].b\mathrel{+}=BranchSet.getCount(i)$}
     \EndFor
    \ForAll{$ID \in V$}
      \State{$V[ID].l\mathrel{/}=V[ID].f$}
      \State{$V[ID].b\mathrel{/}=V[ID].f$}
    \EndFor
  \end{algorithmic}
\end{algorithm}

\textbf{Inter-procedural Analysis:} \prgname takes a summary based approach
to perform the inter-procedural analysis. For a method $M_a$, its semantic vector
$V_a$ is the summary. To perform the inter-procedural analysis, \prgname builds the
summary for each method in the reverse topological order of the call graph. While
processing each method, if line 3 of \Cref{alg:intra} encounters a method invocation, line 4 to line 7 of 
\Cref{alg:intra} will be changed to the process in \Cref{alg:inter}.

\begin{algorithm}
  \caption{Handling Summaries}
\label{alg:inter}
  \begin{algorithmic}[1]
  \Require{A invoke instruction $i$, the semantic vector $V$}
    \State{$V_i \leftarrow Summary(i)$}
    \If{$V_i = null$}
       \State{$ID \leftarrow MAP.getID(i)$}
       \State{$V[ID].f \mathrel{+}= 1$}
       \State{$V[ID].l \mathrel{+}= LoopTree.getDepth(i)$}
      \State{$V[ID].b \mathrel{+}= BranchSet.getCount(i)$}
    \Else
   		\ForAll{$ID \in V.IDset()$}
         \State{$V[ID].f \mathrel{+}= V_i[ID].f$}
         \State{$V[ID].l \mathrel{+}= (LoopTree.getDepth(i)+V_i[ID].l)*V_i[ID].f$}
        \State{$V[ID].b \mathrel{+}= (BranchSet.getCount(i)+V_i[ID].b)*V_i[ID].f$}
      \EndFor
    \EndIf
  \end{algorithmic}
\end{algorithm}

Our summary handling process, which is shown in \Cref{alg:inter}, takes the invocation instruction, $i$, and the semantic vector of the current method, $V$, as an input.  It first detects whether $i$ contains summaries at line 2. If not, it means $i$ points to an API, then \Cref{alg:inter} will process $i$ as a normal instruction and takes the same steps of \Cref{alg:intra}. If the invocation has a summary, suppose it is $V_i$, \Cref{alg:inter} first adds the frequency of all instructions in $V_i$ to the summary of the current method, $V$, at line 9. Then, at line 10, it updates the total loop depths of all instructions of $V$. Since the average loop depth of each instruction in $V_i$ is increased by the loop depth of the instruction $i$ in the current method. The total loop depth in the summary of the current method should be increased by  $(LoopTree.getDepth(i)+V_i[ID].l)*V_i[ID].f$. The similar updates are also performed for the total branch count  at line 11. 

After the process of \Cref{alg:inter} for $i$ has finished, it will return to the
\Cref{alg:intra} to process the next instruction. Finally, the average loop depth and branch count will be calculated at  line 10 and 11 of \Cref{alg:intra}.

\textbf{Processing UI Layout Files:} \prgname takes the UI layout XML files of an Android app as the input and generates the layout vector. The process of this is
similar to the process of processing executables. The only difference is that, for UI
layout XML files, \prgname counts the depth of each UI element in the XML tree rather than the loop depth or branch count.
Android Layout XML files may contain reference tags, which allow people to represent
the layout tree in another file with one XML tag. To handle the reference tags,
\prgname treats them as method calls in the executable and takes the similar
summary based inter-procedural analysis process.

\subsection{The Pre-Training Stage}
\label{sec:preprocessing}
The inputs of the pre-training stage are the semantic vector and the layout vector.
Its outputs are the normalized executable feature and UI feature. The reason of having the pre-training stage is that the semantic vector and the layout vector are not in the same shape and have different magnitude of values. Specifically, in \prgname,
the elements of the semantic vector are 3-tuples while the elements of the layout vector are 2-tuples. They cannot be combined together as one feature vector directly.
Furthermore, the number of instructions can often be more than hundreds of thousands in the semantic vector while the number of UI elements is below one thousand in the layout vector. Any simple method that reshapes and combines the
semantic vector and the layout vector directly will make the machine learning techniques ignore the effect of the layout vector.

\prgname learns the normalized executable feature vector and the layout feature vector with two \ac{nn} models respectively. These two models have similar structures. The
only difference is the shape of the input and the size of hidden layers in the
models.  For the conciseness of this paper, we will focus on the pre-training model of the executable feature.

 The model for executable feature learning is a convolutional neural network model.  The first is a convolutional layer with 10 different $3\times20$ filers. This convolutional layer is followed by three dense layers. Each of the first two layers consists of 1000 nodes and the third layer is with 50 nodes as the feature layer. The last layer of the model is a
 binary classifier to determine whether the input app is a low rating app or not. Each layer uses the $tanh$ as the activation function and utilizes batch normalization to avoid over fitting. Softmax function is used as the classifier and cross entropy loss is used for supervised training.

% \textbf{dongjin comment: what's the form of softmax function and cross-entropy loss? should we include that?}

For training, \prgname first converts the semantic vectors of
 the apps in the training set into a set of $3\times N$ matrices, where N is the total number of types of instructions. Then these matrices are fed to the convolutional layer. During the training, the model will optimize the cross entropy loss to improve the accuracy of classifying low rating apps. After the training, to obtain the executable feature of an app, \prgname feeds the semantic vectors of the app to the pre-trained model and uses the output of the feature layer as the executable feature for the input app.

\textbf{UI Feature: } We can follow the similar procedure to pre-train a CNN model for the semantic feature. The differences are that it contains 10 different $2\times 20$ filters in the convolutional layer and has only one dense layer with 50 nodes as the feature layer.

% dongjin: no hidden layer for UI feature? hidden layer size for UI feature?

\subsection{The Learning Stage}
The input of the learning stage is a vector generated by concatenating the one dimensional executable feature and the one dimensional UI feature. The output is a softmax function to predict whether the input app is a low rating app. The multilayer perceptron used in the learning stage has two dense layers and one output layer for classification. The first dense layer has 100 nodes and the second layer has 20 nodes. The output layer has two nodes to generate the classes of the input app. Class 0 presents the input is a low rating app and class 1 represents that the input app is not a low rating app.
The dense layers use $tanh$ as the activation function. They also use batch normalization to avoid over fitting. The
cross entropy loss is used as the loss function.

\section{Preliminary Study }
\label{sec:preliminary}
The quality of the semantic vector is critical for the accuracy of \prgname. One important question is that \textbf{whether the semantic vector can accurately model the differences and similarities between Java/Android programs on the semantic level}. This question is fundamental because if the semantic vector cannot represent the differences and similarities between Java/Android programs on the semantic level with high accuracy, using it to detect low rating Android apps cannot achieve a good result. Due to this reason, before evaluating the accuracy of using \prgname to detect low rating Android apps, we first answer the question about whether the semantic vector can represent differences and similarities between Java/Android apps in this preliminary study.

To answer this question,  we use the accuracy of classifying Java programs that implement the same functionalities as the metric to measure how well can the semantic vector capture the similarities and differences between Java/Android programs on the semantic level. Our general experiment is as follows. First, we collect a group of Java applications in $K$   categories.  All the programs in the same category, $C_i, 0<=i<=K$, implement the same functionality. Then, we use a convolutional neural network model to classify the category of each Java program. The convolutional neural network model is similar to the pre-training model of the executable representation in \Cref{sec:preprocessing}. However, the model in this preliminary study has smaller hidden layers and the output layer generates $K$ classes instead of two, where $K$ is the total number of categories. Finally, we report the classification accuracy.

The data set of the preliminary study was collected from the
Google Code Jam Website~\cite{codejam}, which is a
programming competition sponsored by Google every year since 2008. Each year Google Code Jam posts programming questions for more than 10,000 competitors. Each question requires competitors to upload a piece of program that can pass the pre-designed test suite with required resource. Thus, the uploaded answers to the same question implement the same functionality and belong to the same category in our preliminary strudy.

In this our preliminary study, we downloaded the Java answers from the top 100 competitors from 2008 to 2016.
We filtered the programs as follows:
\begin{itemize}
   \item We removed those solutions that could not be compiled.
   \item We removed those solutions that caused Soot to crash.
   \item We removed those questions that had less than 20 answers.
 \end{itemize}

The reason for the first two criteria was that our implementation replied on Soot to analyze the binaries of Java programs. For the
third criterion, we did this because we needed sufficient cases in each category to evaluate the accuracy. After the filtering, our set of test cases contained 105 questions and 8,245 answers. 
In other words, we had 8,245 Java programs in 105 categories and each category contained at least 20 programs.

To further evaluate the accuracy of our executable representation, we compared the accuracy of the using the semantic vector with the
baseline representation, which was the 1-dimensional bag of words approach~\cite{Peiravian2013ICTAI,sahs2012EISIC,Shamili2010ICPR,Aafer2013,schmidt2009static,burguera2011crowdroid,enck2009lightweight,zhou2012hey,arp2014drebin}. This representation equals to only using the frequency of instructions in the semantic vector. We built two baseline neural network models that accepted the bag of words representation as the input.
 The first one was the fully connected model. In this model,  we replaced the convolutional layer to a dense layer. The second model was a baseline convolutional neural network model. For this model, we replaced the 3$\times$20 filters in the semantic vector model with the 1$\times$20 filters since the bag of words representation only has one dimension.

During our experiment, we first calculated the semantic vector for each of the
programs with the method introduced in~\Cref{sec:exerepresentation}. Then, we used
the convoluational neural network model to classify the programs. To have a valid result, we took a
10-fold approach and repeated
the experiment for ten times. In each round of the experiment, we randomly split the
data into ten sets. Then, we used nine sets as training sets and one as the testing
set. Then the classification
accuracy was measured in each experiment. Finally, we reported the average results and the standard deviations. The same protocol was applied to the two baseline models for the bag of words representation.

In our measurement, the neural network model on the semantic vectors achieved an average
accuracy of 91.86\% with a standard
deviation of 1.00\%. The fully connected neural network model over the bag of words representation
achieved on average 89.64\% accuracy with a standard
deviation of 1.30\%. The convolutional neural network model over the bag of words representation achieved on average 90.50\% accuracy with a standard deviation of 1.08\%. We compared the result of the semantic vector model to the two baseline models with student test. The p values were below 0.04, which meant that the accuracy of the model on the semantic
vectors was significantly higher than the two baseline models.

This result suggests that the semantic vectors can accurately represent the semantic
similarities between programs. It indicates that the semantic vector can possibly capture the semantic features of a Java/Android program. It is possible to use the semantic vector to detect low rating apps.  

Our result also shows that the semantic vector has a significantly higher accuracy
than the bag of words representation. This is not surprising since the semantic vector of
a program contains the loop and branch information, which is not contained by the bag of words representation.

% One interesting fact in our experiment is that even the simple bag of word
% representation can achieve a 90.50\% of accuracy with the \ac{nn} model. This is because the test cases in our preliminary study are small Java programs with hundreds of lines. For these applications, omitting the loop and branch information may not lose too much semantic level information. However, with larger and more complex apps, neglecting loops can branches can be more problematic. Nevertheless, as we evaluated, even with this small Java applications, the semantic vector still outperforms the baseline bag of word approach.

\section{Evaluation}
\label{sec:evaluation}
In our evaluation,  we seek to answer the following three research questions:
\begin{itemize}
  \item RQ 1: How accurate can our approach detect low rating Android apps
  \item RQ 2: Processing time of static analysis
  \item RQ 3: The learning time
\end{itemize}

\subsection{Implementation}

We implemented the analysis stage of \prgname with soot~\cite{soot},
apktool~\cite{apktool}, and FlowDroid~\cite{Arzt:2014:FPC:2666356.2594299}. We used
apktool to unpack Android .apk files and retrieve the XML files.
We used FlowDroid to build the call graph of Android apps. We used soot to build the
semantic vector of the executable.
We used Keras~\cite{keras} and tensor flow~\cite{tensorflow} to implement the neural network
models. We used the API set Android 7.0 as the instruction set of the semantic vector. We also used the UI element set of Android 7.0 for the layout vector.
The hardware we used in our experiments was a desktop with Core i7 6700K processor,
32GB memory, and Nvidia GTX 1080 graphic card.
In our implementation, we trained the model with batched input. The batch size was 128. We also trained the training data for ten epochs to achieve the best accuracy.

\subsection{Data Set}
We evaluated \prgname with real Google Play apps. In our experiments, we downloaded
Google Play apps from the PlayDrone
project~\cite{playDrone,Viennot:2014:MSG:2591971.2592003}, which contains 1.1 million
free apps and their meta information from the  Google Play market. In our experiment,
we categorized the apps based on their star numbers into four groups: 1 to 2
stars, 2 to 3 stars, 3 to 4 stars, and 4 to 5 stars. For each category, we downloaded
10,000 apps randomly. Thus, we had  40,000 apps from PlayDrone and the stars of
these apps are roughly evenly distributed from one star to five stars. For these
downloaded apps, we filtered out the apps that caused soot or FlowDroid to crash
and the apps that their UI layout could not be successfully retrieved by apktool.
After the filtering,
we had 8,684 apps with more than four stars, 7,854 apps with
three to four stars, 8,775 apps with two to three stars, and 8,142 apps with less than two starts. In total, we had 33,456 market Android apps.

Our apps were from 25
categories, such as game, lifestyle, and business. \Cref{fig:Piechart} shows the distribution of categories
of our downloaded apps.
\oneColFig{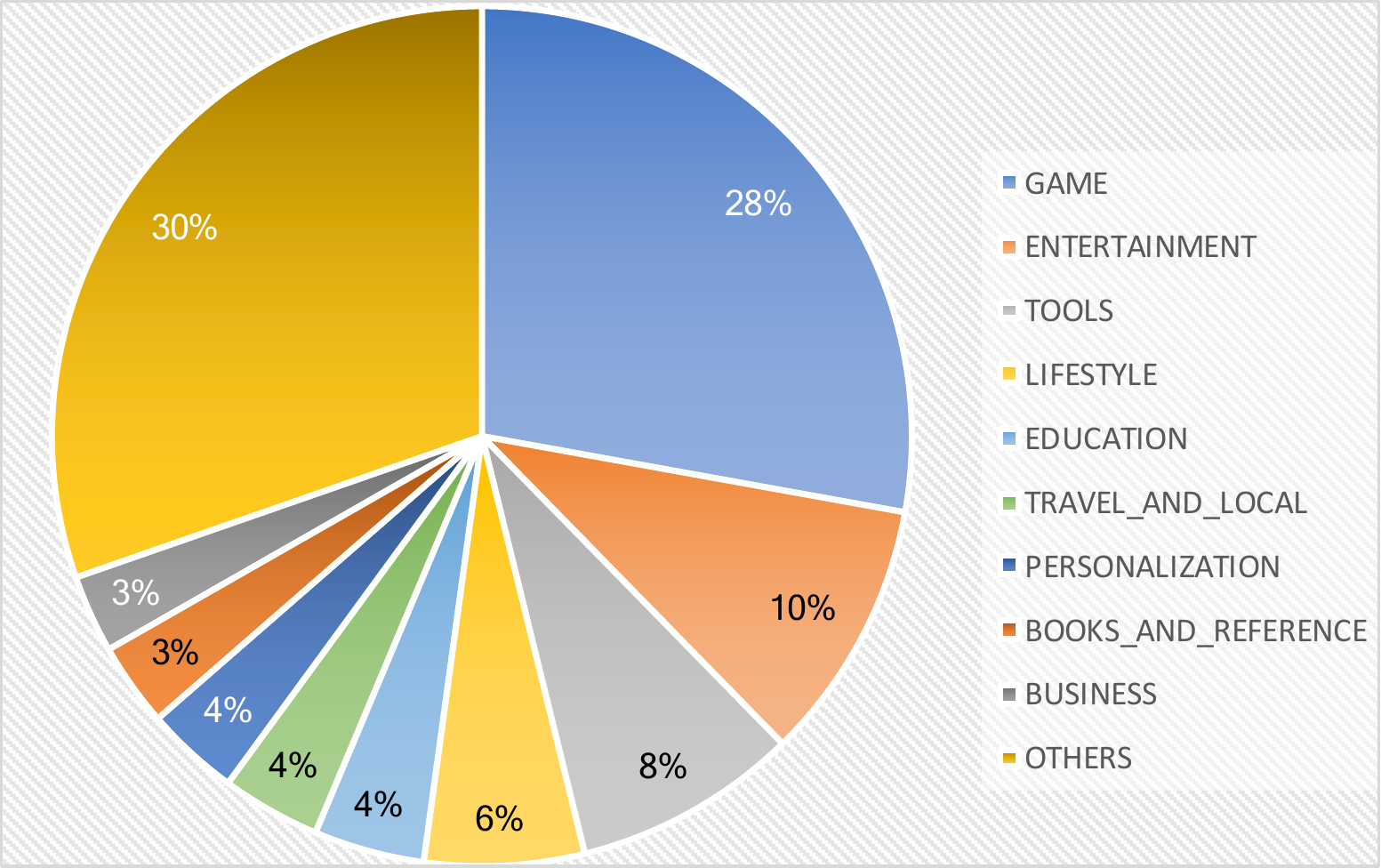}{The distribution of categories of our apps}
In~\Cref{fig:Piechart}, we plotted the top 9 categories with the most apps as the
pie chart. All other 16 categories were summarized as OTHERS. As shown in the chart,
we had a significant amount of apps in each category. The GAME category had the
most apps because there were much more games than other apps in the Google
Play~\cite{Viennot:2014:MSG:2591971.2592003}. In fact, GAME could be further broken
down into 17 sub categories.  Nevertheless, our downloaded apps could still provide enough diversity in the functionalities of apps.

\subsection{RQ 1: Accuracy of Classification}
\label{sec:rq1}
Our first research question is to evaluate the accuracy of \prgname of detecting low rating apps.
To do this, we first labeled all the apps that have less than three stars as low rating apps. For all the apps that have more than three stars, we labeled them as negative samples. Thus, in total, we labeled 16,917 apps as low rating apps and 16,538 apps as not-low rating apps. Then we followed the 10 fold protocol we had used in
\Cref{sec:preliminary}: we randomly split the data set into ten parts, took nine of
them as the training set and one as the testing set. This process was repeated for
ten times. The averages and standard deviations of precisions, recalls, and accuracies were measured.

In our evaluation, we also built four other baseline models for comparison.  The first two were based on the bag of words representation,  which was used in previous techniques~\cite{Peiravian2013ICTAI,sahs2012EISIC,Shamili2010ICPR,Aafer2013,schmidt2009static,burguera2011crowdroid,enck2009lightweight,zhou2012hey,arp2014drebin}. Our first model was the fully connected neural network model for the bag of words representation. The structure of this model was similar to the re-training model of the semantic vector. The only difference was that we replaced the convolutional layer of the pre-training model of the semantic vector with a dense layer with 1000 nodes.  
The second model was the \ac{nn} model over the bag of words representation.  In this model, we replaced the 3$\times$20 filters in the pre-training model for the semantic vector with 1$\times$20 filters to accept the 1-dimensional bag of words representation. The third model we built was an executable only model. In
this model, we used the pre-training model for the semantic vector learning alone to detect low rating apps. Similarly, our fourth model was the UI only model. We built it by using
the pre-training model
for UI feature learning alone. This model evaluated the accuracy of only using UI
information for
low rating app detection. For all these four models, we followed the same 10 fold experiment
process and calculated the averages and standard deviations of their precisions, recalls, and accuracies.

\oneColFig{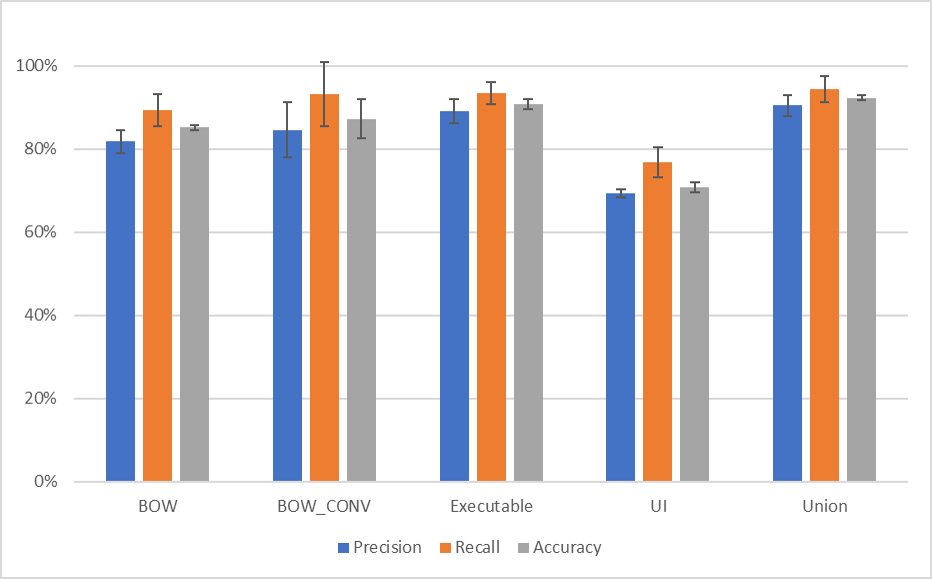}{The accuracy of \prgname}

The result of our measurement is shown in~\Cref{fig:accuracy}. The bars with $BOW$
represent the result of the bag of words representation with the fully connected neural network model. $BOW\_CONV$ represents the bag of words representation with the convolutional neural network model. Executable is the result of only
using the semantic vector for classification. UI is the accuracy of only using the
UI representation. Union is the accuracy of \prgname.

In our experiment,
\prgname on average achieved 92.31\% of accuracy with the standard deviation as
0.55\%. The average precision was 90.50\% with a standard deviation of 2.25\%. The average recall was 94.31\% with a standard deviation of 3.14\%.

For the bag of words representation with the fully connected neural network model, it achieved an accuracy of 85.14\% with a standard
deviation of 0.67\%. Its average precision was 81.76\% with a standard deviation of 2.78\%. Its average recall was 89.32\% with a standard deviation of 3.77\%. 

For the bag of words representation with the convolutional neural network model,
the average accuracy of was 87.18\%
and the standard deviation was 4.7\%.  The average precision was 84.63\% with a standard deviation of 6.7\%. The average recall was 93.12\% with a standard deviation of 7.70\%. 

For the executable only model, the average accuracy was 90.82\%
and the standard deviation was 1.25\%. The average precision  was 89.10\% with a standard deviation of 2.94\%. The average recall of the executable only model was 93.49\% with a standard deviation of 2.68\%.

For the the UI only model, the average accuracy
was 70.73\% with a standard deviation of 1.15\%. The average precision  was 69.29 with a standard deviation of 0.92\%. The average recall was 76.75\% with a standard deviation of 3.62\%. 

To ensure the statistical significance, we also made a student test between the results of each pair of the five models. The p-value of all tests were smaller than
0.035, except the recall between the $BOW\_CONV$ and $Executable$. This result meant that there were statistically significant differences between the results of the five models  except the recalls of $BOW\_CONV$ and $Executable$.

The result of our experiment is promising. It shows that \prgname can accurately detect low rating apps. Note that, in our implementation, we simply labeled apps with less than three stars as low rating apps. This labeling method does not consider the borderline apps which have stars around three. For example, an app with 2.9 stars is not necessarily worse or less popular than an app with 3.1 stars. It is harder for \prgname to correctly detect the borderline apps. Due to this reason, we think the 92.31\% of accuracy is satisfiable.

One interesting observation for our result is that \prgname has a higher recall than precision. This result means that our approach is less likely to miss low rating apps. This result is more beneficial for app stores because they can use \prgname to scan their apps and find out all candidates of low rating apps. This process will only miss 5.7\% of low rating apps. Then, the app stores can focus on the apps that are detected as low rating apps. This process could potentially save the labor for app markets to detect low rating apps.

The results of our experiments also prove that it is effective to add branch and
loop information into the semantic vector. As shown in~\Cref{fig:accuracy}, using
the semantic vector can achieve 3.62\% more accuracy than the bag of words approach. It can also achieve a much smaller standard deviation, which means the model with the semantic vector is more robust than the convolutional neural network model over the bag of word representation. The improvement of using the semantic vector to detect low rating apps is larger than the improvement of using it to classify programs with the same functionality in~\Cref{sec:preliminary}. This is because realistic Android apps are larger and more complex than solutions to programming questions on the CodeJam platform.  It is more beneficial to keep the loop and branch information in large programs. 

Our experiments also prove that combining the UI representation
with the executable representation can significantly improve the detection
accuracy over the executable only model. This is not surprising since UI is also an important factor that can influence user experience of an Android app.

\subsection{RQ 2: Run Time Overhead of Static Analysis}
Our second research question is to evaluate the time consumed by our
static analysis stage to build the executable representation and UI
representation from the .apk file of an Android app. 

To answer this research question,  we first added time stamp logs to our static analysis code and scripts.
Then we executed our code and collected the time overhead for each of our test cases.
To have a better understanding of the time overhead of the static analysis phase, we broke down the total time as four categories: the processing time of FlowDroid, the time overhead to build the
executable representation, the time for apktool to retrieve the UI xml files, and
the analysis time to build the UI representation. 

In our measurement, the average processing time was 23.98 seconds per app. The standard deviation was 18.00 seconds. The breakdown of the four categories of time
is shown in~\Cref{fig:TimeBreakDown}. More specifically, on average, the
processing time of FlowDroid per each app was 17.44 seconds with a standard
deviation of 17.36 seconds, the time overhead to build the executable
representation was 1.22 seconds and the standard deviation was 0.77 seconds, The
 the time cost of apktool to retrieve UI xml files was 5.14 seconds with a standard deviation of 0.42 seconds, and the processing time to build the UI representation was 0.28 seconds with a standard deviation of 0.04 seconds.

\oneColFig{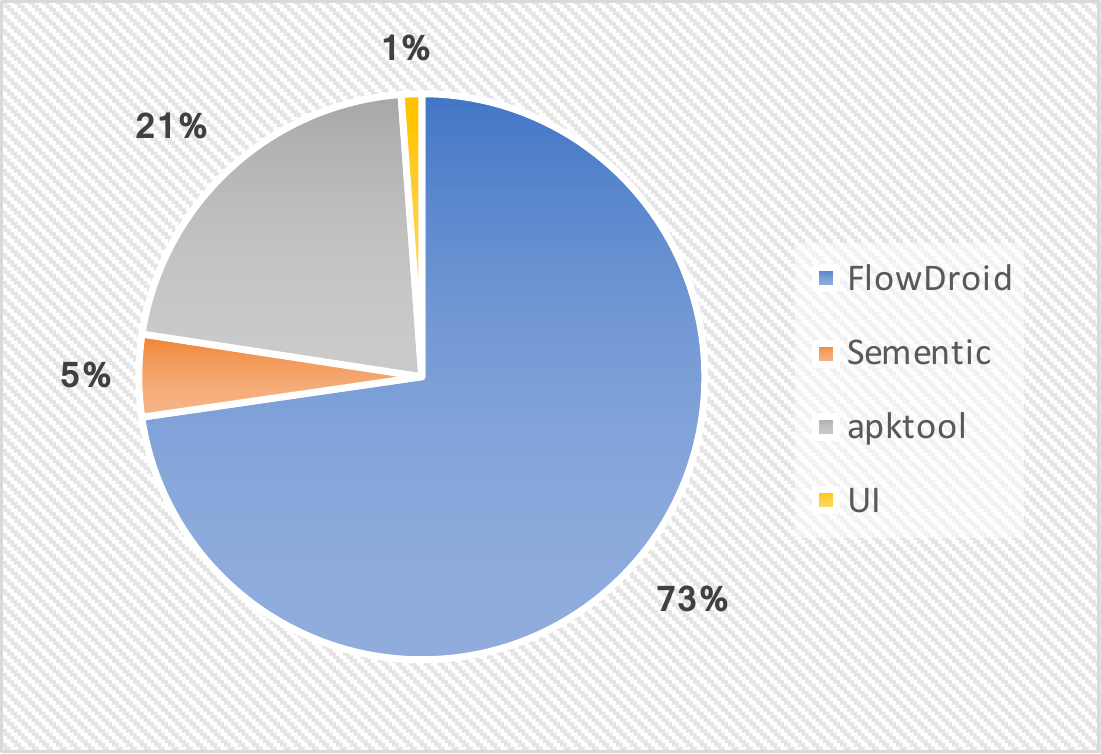}{The time cost of the static analysis of \prgname}

According to the result, our static analysis stage on average take less than 30
seconds to process one Android market app. This indicates that our approach is very
scalable and can be applied to realistic Android markets. One interesting fact
about our result is that 94\% of the processing time of our approach is consumed by
the apktool and FlowDroid. Especially, 73\% of the time is consumed by FlowDroid.
This is because besides building the call graph, FlowDroid also does other
analyses which are not used in our approach.  We expect our
approach could be faster if we use a more light-weighted way to build the call
graph.

\subsection{RQ 3: Learning Time}
To answer this research question, we measured the time consumed during the training
stage. This time contains three parts: the time consumed by the pre-training model of the
executable feature learning, the time of the pre-training model of the UI feature learning, and the time of the final learning stage. To ensure the validity of our result, we measured the time consumption of each of the three parts of the learning time for ten times during the 10 fold experiment.
\begin{figure}[htbp]
\centering
\includegraphics[width=0.45\textwidth]{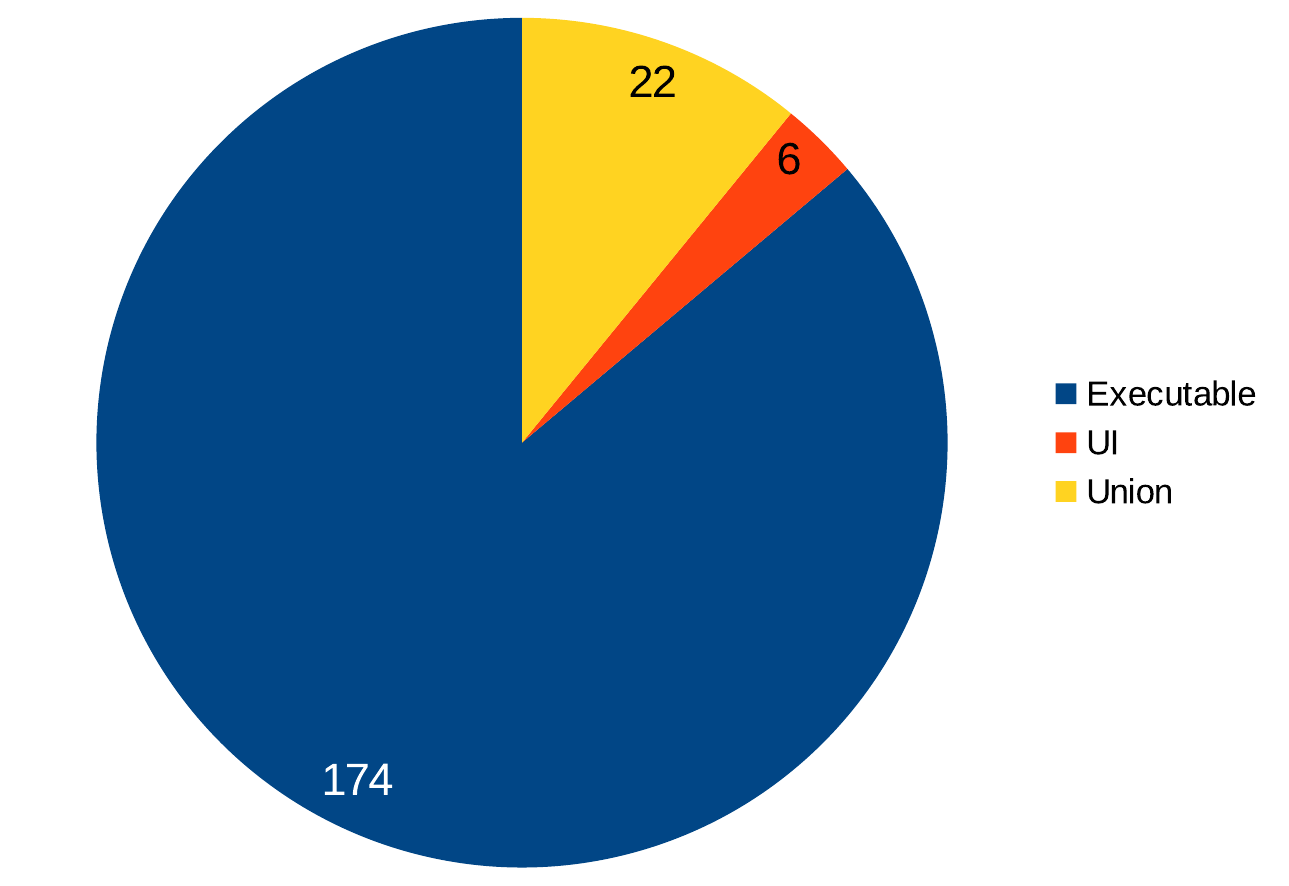}
%\vspace{3pt}
\caption{The learning time of \prgname}
\label{fig:LearningTime}
\end{figure}
 The result is shown in \Cref{fig:LearningTime}, where the unit is minute. On average,
 the pre-training model for the executable feature was 174 minutes with a standard deviation of 28 minutes. The average time to train the UI feature model was six minutes with a standard deviation of 10 seconds. The average training time in the final stage was 22 minutes with a standard deviation of 17 seconds.

 As shown in our evaluation, our model could be trained in 3.4 hours. It is acceptable for more than 30,000 apps.  This time can be further reduced with more powerful hardware. Another fact in our experiment is that  86\% of the total training time is
 spent while training the executable feature model in the pre-training stage. This because the input of the semantic vector has much larger dimensions than the UI representation and the input in the final stage. Nevertheless, our model can still be trained in a reasonable amount of time.

\section{Threat To Validity}
\label{sec:threat}
\textbf{External Validity:} To guarantee that our applications are representative, we collected 33,456 realistic Android apps from
the Google Play market. These apps have different star ratings and are from 25 categories. During the process of app collection,
we filtered the apps that could not be processed by soot and FlowDoird and the apps whose layout files could not be retrieved by
apktool.  This process did not bias the result of our evaluation.

In our experiment, we used the apps from the database crawled by PlayDrone~\cite{playDrone,Viennot:2014:MSG:2591971.2592003}, it contains the
snapshot of the Google Play store at Oct 31st, 2014. This app set does not contain the latest Android apps since Android 4.4. This threat
does not affect the validity of our experiment because the architecture, API set, and the way to create apps for Android have not
been changed significantly since Android 4.0. To further alleviate
this threat, in our approach, we used the API set and the UI element set of Android 7.0 in our implementation.  The unique APIs and UI elements in Android 4.4 were not captured. Thus, we do not expect the result of our experiments will be significantly different with recent apps.

Our app set only contains free apps. However, we believe that there is not a significant difference between free apps and paid apps in the methodology of programming and UI designing. Thus, we do not expect a very different result of our experiment in paid apps.

\textbf{Internal Validity:}
The neural network models of \prgname were randomly initialized. Such a randomness can affect the accuracy of our models.
The time measurement in our experiments can also be affected by the randomness.
To alleviate the impact of the randomness, we followed 10 fold process. We randomly divided our apps into ten sets. In each
experiment, we used nine sets for training and one for testing. We then averaged the results in ten experiments and reported the
standard deviation.

\textbf{Construction Validity:}
In \Cref{sec:rq1}, we compared the accuracy of \prgname to four baseline models. Such a comparison can be affected by the random measurements errors. To make the result comparison solid, we also performed the student test for each pair of our models. The result of our student test showed that the difference in accuracy between any two models in  \Cref{sec:rq1} was statistically significant.

\section{Related Work}
\label{sec:related}
To the best of our knowledge, detecting low rating  Android apps with
only static program and UI information is a problem that has not been addressed
before. Monett and colleagues proposed a technique to predict the star number of Android apps from the user reviews~\cite{7733272}.  Similar technique is also
used to predict movie scores or other sentiment analysis in
natural language processing~\cite{pang2008opinion,Pang:2002:TUS:1118693.1118704,Pang:2005:SSE:1219840.1219855,Goldberg:2006:SSA:1654758.1654769,Tang:2015:UMN:2832415.2832436}. The problem of
these techniques is that they still require users to provide reviews for Android
apps. It suffers the same problem of the current star rating system.

Another related work to us is Mou and colleagues' work~\cite{Mou2016AAAI}.  This
approach
 first embeds the keywords of a programming language as vectors so that
similar keywords have closer vectors in the Euclidean space. Then, it uses the
token level embeddings to classify the programs from the online judgment platform
of the Peking University.
To classify the programs, this approach represents the nodes in
the \ac{ast} of a program with the token level embeddings  and  uses a tree-based
convolution neural networks to classify the \acp{ast}. This task is similar to what we have done in the preliminary study.

Compared to Mou and colleague's work, our approach addresses a different problem with realistic apps. Our approach detects the
low rating Android market apps. Android apps are substantially larger and more complex than online judgment
platform questions. The
star ratings of Android apps are also less accurate than the labels of questions.
Furthermore, Mou and colleague's work requires the source code to build the
\acp{ast}, while our approach works directly on the executables. Thus, our approach
can be used on close-source applications while Mou and colleague's work cannot.

White and colleagues  proposed a neural network based
techniques to detect program clones~\cite{White2016ASE}. This work first uses
recursive neural network based approach  to embed tokens of program source code as
vectors~\cite{White2015TDL}. It then
encodes the source code of program snippets as vectors with a recursive
autoencoder~\cite{socher2011semi}. Finally, the program vectors are used to detect
program clones.  Deckard~\cite{jiang2007deckard,gabel2008scalable} summarizes the patterns in \acp{ast} of programs and counts these
 patterns as vectors. Then it uses machine learning techniques to detect code
 clones with the tree patterns.  Chen and colleagues embeds \ac{cfg} of programs as
 a matrix~\cite{chen2014achieving} to detect program clones. Unlike \prgname, these
 approaches focus on detecting program clones. They do not detect low rating Android apps. Further, similar to Mou and colleague's
work~\cite{Mou2016AAAI}, these approaches also
need the source code, while our approach does not.

Mu and colleagues proposed a machine learning based technique, DroidSIFT,  to detect
malwares~\cite{Zhang:2014:SAM:2660267.2660359}. This approach builds the API
dependency pattern database from benign and malicious Android apps. Then it encodes
new apps based on its similarities to the API dependency patterns in the databases.
Finally, it detects malwares by learning a model from the app encodings. The
limitation of this work is that it encodes apps based on the dependency graph for
each API. It is very expensive to build the dependency graph for all APIs used by an app. For malware classification, this problem can be alleviated by focusing on a
small set of permission related APIs. However, for detecting low rating apps, people cannot only analyze a small set of
APIs. In this case, DroidSIFT will encounter the scalability issue.

Wang and colleagues proposed a neural network based approach for defect
prediction on the source code file level~\cite{Wang:2016:ALS:2884781.2884804}.  This approach first encodes
programs based on the token vector of AST nodes. Then it uses a deep belief network
for feature dimension reduction. Finally, it trains a model to classify buggy
files with the features with reduced dimension. This technique cannot be directly used
to detect low rating apps because of two reasons. First, it requires source code. Second,
the size of the token vector can be too large for machine learning models on the whole application level.

Bag of word approaches are used for malware
detection~\cite{Peiravian2013ICTAI,sahs2012EISIC,Shamili2010ICPR,Aafer2013,schmidt2009static,burguera2011crowdroid,enck2009lightweight,zhou2012hey,arp2014drebin}.
Compared with these approaches, \prgname addresses a very different problem.
Furthermore, as we evaluated in \Cref{sec:rq1}, our approach significantly
outperforms the bag of word approach regarding classification accuracy.

Many approaches also use machine learning techniques to generate code snippets from
natural language
queries~\cite{Gu:2016:DAL:2950290.2950334,Allamanis:2015:BMS:3045118.3045344,Raghothaman:2016:SSI:2884781.2884808}. These techniques learn a translation model from
the API sequences and their comments. Then, when people provide the model a
natural language query, the model will generate the API sequence. Other techniques
are also proposed for API patterns
mining~\cite{Fowkes:2016:PPA:2950290.2950319,6693127,Wang:2013:MSH:2487085.2487146}.
Despite the usefulness of these techniques, they address very different problems as
we addressed in this paper.

There are also a group of studies that examine the relationship between the
rating and features of apps. Tian and colleagues~\cite{7332476} found that
the size, code complexity, and other 15 features have positive correlations with
the star ratings. Linares-Vasquez and
colleagues~\cite{Linares-Vasquez:2013:ACF:2491411.2491428} studied the
relationships of API changes in Android apps and star ratings. Ruiz and
colleagues~\cite{6811104} studied the correlation between ad libraries and ratings
of Android apps. Gui and colleagues~\cite{7194565} studied the relationship between
energy consumption and ratings of Android apps. Although the conclusions of these
techniques are interesting, they do not have a method to predict or classify
low rating apps.

\section{Conclusions}
\label{sec:conclusion}

In this paper, we proposed a novel method to detect low rating Android apps only based on the .apk files. With our approach, an app market does not need to risk its reputation by exposing low rating apps to end users for rating collection. Our approach is based on static program analysis and machine learning. In our approach, we first proposed novel representations for the executable file as well as the UI layout of an Android app.  Then, based on the executable and UI representations, we built a convolutional neural network model to detect low rating Android apps.
 
We also performed an extensive evaluation of our approach on 33,456 realistic Android market apps. In our experiment, our approach could detect low rating Android apps with 90.50\% precision and 94.32\% recall. Our approach also outperformed all the four baseline models with statistical significance.
 
Overall, our approach is both accurate and scalable. It can be potentially helpful for Android app markets to prevent low rating apps and accelerate manual reviewing process.

%\input{Acknowledgement}
%!TEX root = paper.tex

%\clearpage
\balance
{
\bibliographystyle{ACM-Reference-Format}

%\footnotesize 
\bibliography{paper}
}

\end{document}